# A Novel Fault Classification Scheme Based on Least Square SVM

H.C. Dubey[1], A.K. Tiwari[2], Nandita[3], P.K. Ray[4], S.R. Mohanty[2] and Nand Kishor[2]

*Abstract*-- This paper presents a novel approach for fault classification and section identification in a series compensated transmission line based on least square support vector machine. The current signal corresponding to one-fourth of the post fault cycle is used as input to proposed modular LS-SVM classifier. The proposed scheme uses four binary classifier; three for selection of three phases and fourth for ground detection. The proposed classification scheme is found to be accurate and reliable in presence of noise as well. The simulation results validate the efficacy of proposed scheme for accurate classification of fault in a series compensated transmission line.

*Index Terms*-- Digital relaying, distance relay, fault classification, least square-support vector machine, modular classifier, series compensated transmission line.

## I. INTRODUCTION

Transmission lines are protected by digital relays which operate accurately and reliably as compared to their solid state counterparts. Digital relay are invariably based on signal processing techniques. Faults occurring in transmission lines need to be detected, classified, and located fastly and accurately for keeping faulty part isolated from the healthy part thereby, ensuring minimum damage and disturbance. The speed, accuracy and reliability of digital protective relays are dependent on underlying techniques used in design. Fault classification task is an important task after the detection of fault for isolating the faulty section from healthy one, thus reducing damages to power systems [1]-[2].

The FACTS employed in modern power system achieve enhancement in transmittable power, improvement in the system stability, reduction in transmission losses and more flexibility in power control. On the other hand, presence of FACTS devices like TCSC complicates the protection task because [1, 2]: the metal oxide varistor (MOV) is employed for protection of the series capacitor from over-voltage during faulty situations. However, it acts non-linearly during faults and increases the complexity of the protection problem. Due to resonance between the system inductance and series capacitor, non fundamental decaying components as well as decaying DC components are present in the voltage and current signals. Odd harmonics are due to MOV conduction during faults and sub-synchronous frequencies having frequency components varying around half the fundamental frequency value, high frequency components (which results from resonance between line capacitance and line inductance) are present in addition to fundamental components of the steady state fault current. Thus, these signals are not processed very accurately with conventional methods such as full cycle DFT/half cycle DFT (FCDFT/HCDFT) or least square error (LSE) technique; thereby cause large error in estimation of the fundamental phasor. In light of these issues, the fault classification task is complicated in the presence of TCSC [3]-[4]. Techniques based on fuzzy logic, adaptive Kalman filtering, neural network, wavelet transform and support vector machine have been addressed in research studies for fault classification task [5-14]. In [5], Kalman filtering based technique has been used for fault classification task. However, the validity of the proposed method had been tested on a limited number of test cases. Classification technique based on artificial neural network (ANN) in a series compensated transmission lines is addressed in [6]-[8]. However, ANN involves empirical risk minimization and suffers from drawbacks like over fitting and trapping in local minima. Further in [6]-[8], the performance of the classification schemes have not been investigated over extensive test cases and the compensation level was kept constant.

In [9]-[10], authors have suggested classification schemes based on wavelet transformation. Also, in [9]-[10], the proposed technique was validated only for limited cases. Some techniques based on fuzzy logic have also been presented for fault classification applications in [11]-[13]. A hybrid approach based on wavelet transform and fuzzy logic is proposed in [11]-[12]. Combination of fuzzy logic and higher order statistics (HOS) has been investigated in [13] for classification task considering a wide variation in the system conditions. Recently, support vector machine (SVM) based technique has been proposed for fault classification [14]-[17].

[1]Harish Chandra Dubey is with the Department of Electronics and Communication Engineering, Motilal Nehru National Institute of Technology, Allahabad-211004, INDIA (e-mail: dubeyhc@ieee.org ).
[2]Ashutosh Kumar Tiwari, Soumya Ranjan Mohanty and Nand Kishor are with the Department of Electrical Engineering, Motilal Nehru National Institute of Technology, Allahabad-211004, INDIA(e-mail: aktiwari@ieee.org , soumya@mnnit.ac.in, nandkishor@mnnit.ac.in ).
[3]Nandita received the Bachelor of Technology degree in Electronics and Communication Engineering from Birla Institute of Technology (BIT), Mesra-814142, INDIA (email: nandita16bit@gmail.com).
[4]Prakash Kumar Ray is with the Department of Electrical and Electronics Engineering, International Institute of Information Technology, Bhubaneswar- 751003, INDIA (email: prakash@iiit-bh.ac.in).

The presence of FACTS devices in transmission line influence pre-fault as well as post-fault current signals thereby making the task of fault classification a complicated one. In spite of research studies done, fault classification in series compensated transmission lines still remains a challenge in terms of accurate and reliable classification with reduced computational burden and number of data sets for sake of online implementation. One of the most promising approach used for fault classification applications in a transmission line as reported [13, 15] is SVM. SVMs are classifiers based on structural risk minimization. An improved version of SVM is least square SVM (LSSVM) that retains all advantages of SVM with lesser computational burden.

This paper proposes a novel scheme for accurate classification of faults in a series compensated transmission line based on LSSVM classifier using modular topology. The use of modular topology reduces the computational burden and time complexity of the proposed classification technique. The novelty of the work is three fold. First, the fault classification is studied in a series compensated transmission line rather a normal transmission line. Second, the introduction of LSSVM is new to fault classification task. Third, the modular topology of binary LSSVM classifier improves the performance. The proposed classifier is trained with input and output data-sets generated using Simulink model of studied power system in MATLAB environment.

The rest of the paper is presented as follows; the LSSVM is briefly described in section II, proposed scheme is discussed in section III, followed by simulation and results in section IV. Finally, conclusions based on simulation results are given in section V.

## II. LEAST SQUARE SUPPORT VECTOR MACHINE

Support vector machines (SVMs) are statistical learning systems based on structural risk minimization used for pattern recognition. But, the computational complexity of SVMs is related to the number of training samples. As the sample size increases, the solution of the corresponding quadratic programming problem becomes complex resulting in slow computing speed. Least square SVM (LSSVM) is an improvement over SVM without losing its advantages [18]-[19]. LSSVM technique for classification and non-linear regression are characterized by linear Karaush-Kuhn-Tucker (KKT) systems [19]. Sparseness can be imposed in LSSVM using pruning as known in NNs. Non-linear SVM for classification and regression problems involves convex quadratic programming. The LSSVM as a classifier differs from conventional SVM in two ways; firstly it uses equality constraints and secondly, a squared loss function is taken as error variable [19].

Given a training data set of N points $(x_k, y_k)$, k=1,2,…,n where $x_k \in R^n$ is the k-th input data-set and $y_k \in R$ is the k-th output data-set. The SVM based classifier involves a decision function which is given by $y = w^T x + b$. Using the kernel function, the non-linear SVM becomes:

$$y(x) = \sum_{k=1}^{N} \alpha_k K_f(x, x_k) + b \quad (1)$$

The optimal separation of hyperplane is shown in Fig.1. LSSVM aims to build a classifier which is chracterized by following optimization problem:

$$\min_{w,b,e} J_{LS}(w,b,e) = \frac{1}{2} w^T w + \gamma \frac{1}{2} \sum_{k=1}^{N} e_k^2 \quad (2)$$

Subject to the equality constraints

$$y_k = w^T x_k + b + e_k, k = 1\cdots;2,…,n \quad (3)$$

Where $e_k$ are slack variable and $\gamma$ is a positive real constant known as regularization parameter. The parameter $\gamma > 0$ determines the trade-off between fitting error minimization and smoothness.

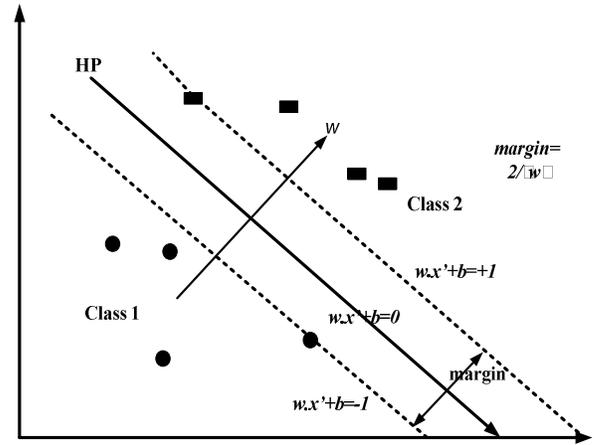

Fig.1 The optimal hyperplane of the binary classifier

The lagrangian involved in the optimization problem is

$$L(w,b,e,a) = w^T w + \gamma \frac{1}{2} \sum_{i=1}^{n} e_i^2 - \sum_{i=1}^{n} a_i (w^T \phi(x_i) + b + e_i - y_i) \quad (4)$$

Where $a_i \in R^N, i = 1,2,…,n$ are Lagrange multipliers. The conditions for optimality are given by:

$$\begin{cases} \frac{\partial L}{\partial w} = 0, gives; w = \sum_{i=1}^{n} a_i \phi(x_i) \\ \frac{\partial L}{\partial b} = 0, gives; \sum_{i=1}^{n} a_i = 0 \\ \frac{\partial L}{\partial b} = 0, gives; a_i = \gamma e_i, i = 1,2,3,…,n \\ \frac{\partial L}{\partial e_i} = 0, gives; b = y_i - w^T \phi(x_i) - e_i; i = 1,2,3,…,n \\ \frac{\partial L}{\partial a_i} = 0 \end{cases} \quad (5)$$

The solution of eqn. (5) is obatined as:

$$\begin{pmatrix} 0 & \vec{1}^T \\ \vec{1} & \Omega+\gamma^{-1}1 \end{pmatrix} \begin{pmatrix} b \\ a \end{pmatrix} = \begin{pmatrix} 0 \\ y \end{pmatrix} \quad (6)$$

$$\vec{1} = [1,1,1,...,1]^T, \quad a = [a_1, a_2, a_3, ..., a_n] \quad (7)$$

$$\Omega_{ij} = \varphi(x_i)^T \varphi(x_j) = K_f(x_i, x_j); i,j = 1,2,3,...,n \quad (8)$$

The classifier based on LS-SVM model is then constructed as follows:

$$\min_{a,b,e} \frac{1}{2} w^T w + \gamma \frac{1}{2} \sum_{i=1}^{n} e_i^2$$
$$such \ that \ e_i = y_i - (w^T \phi(x_i) + b), \quad (9)$$
$$i = 1,2,3,...,n$$

The feature vector $\phi(x)$ is known by means of the positive definite kernel function. The kernel function are required to satisfy Mercer's condition, which implies that

$$K_f(x_k, x_j) = \Phi(x_k)^T \Phi(x_j) \quad (10)$$

The kernel function which are commonly used are linear, polynomial radial basis function (RBF) and multi-layer perceptron (MLP). The selection of regularization parameter $\gamma$ and the kernel parameter $\sigma^2$ is very important for the classifiers. This work uses grid-search to decide appropriate regularization parameter $\gamma$ and the kernel parameter $\sigma^2$. Grid-search is easy and direct as each pair-parameter J,V is independent and it can be sorted in parallel. This classifier finds the optimal parameters in the objective function with cross-validation until the best parameters with high precision are found.

### III. PROPOSED SCHEME

This section presents the application LSSVM to develop the modular multiclass classifier [20]. The block diagram of proposed scheme for fault classification and section identification is shown in Fig.2. The LSSVM based classifier are found to perform better when trained on lesser data sets thus enabling better accuracy with lesser training data sets [18]-[19].

TABLE I

FAULT TYPE IDENTIFICATION

| S.No. | Output of LS SVM$_{-R}$ | Output of LS-SVM$_{-Y}$ | Output of LS-SVM$_{-B}$ | Output of LS-SVM$_{-G}$ | Type of fault |
|---|---|---|---|---|---|
| 1 | 1 | -1 | -1 | 1 | R-G |
| 2 | -1 | 1 | -1 | 1 | Y-G |
| 3 | -1 | -1 | 1 | 1 | B-G |
| 4 | 1 | 1 | -1 | 1 | R-Y-G |
| 5 | 1 | -1 | 1 | 1 | R-B-G |
| 6 | -1 | 1 | 1 | 1 | Y-B-G |
| 7 | 1 | 1 | 1 | 1 | R-Y- B-G |
| 8 | 1 | 1 | -1 | -1 | R-Y |
| 9 | 1 | -1 | 1 | -1 | R-B |
| 10 | -1 | 1 | 1 | -1 | Y-B |

The outputs of four modules of binary classifiers are used for deciding fault type as according to Table I. Here, +1 represents the involvement of phase or ground in fault and -1 represents its absence. The binary LSSVM classifier are combined to form a modular network of four binary classifier, thus forming a multi-class classifier to discriminate between R-G, RY, RY-G, RYB and RYB-G faults.

#### A. LSSVM for section identification

Identification of faulty section is also necessary with fault classification in a series compensated transmission line. Due to presence of FACTS devices the apparent impedance of the transmission changes and hence just after the midpoint and just before that there is a significant change in current, so the section having fault must be identified.

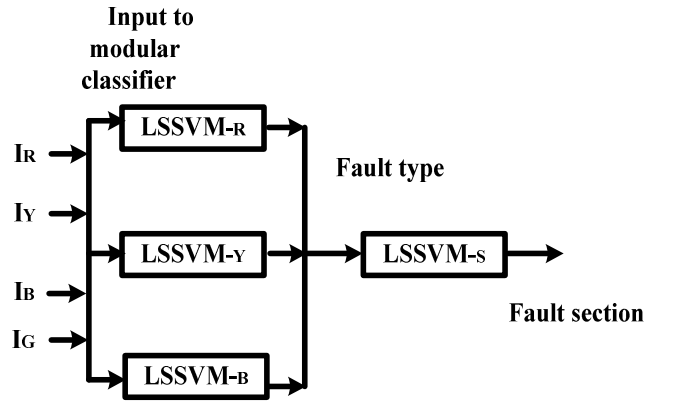

Fig.2 Proposed LSSVM based classification scheme

TABLE II

TESTING OF LSSVM$_{-S}$ FOR SECTION IDENTIFICATION

| Fault | Kernel | Parameter Value | Classification |
|---|---|---|---|
| RY fault at 25%, FIA=135º, LA=40º, R$_f$= 60 ohm | Poly | 1.3269 | 1 |
|  | RBF | 0.208 | 1 |
| RB-G fault at 30%, FIA=175º , LA=30º, R$_f$= 80 ohm | Poly | 1.322 | 1 |
|  | RBF | 0.206 | 1 |
| RYB-G fault at 75%,FIA=120º , A=65º, R$_f$ = 80 ohm | Poly | 1.440 | 1 |
|  | RBF | 0.220 | -1 |
| YB-G fault at 90%, FIA= 95 º LA=55º, Rf= 100 ohm | Poly | 1.389 | 1 |
|  | RBF | 0.276 | -1 |
| RB fault at 30%, FIA=110° LA=45°, $R_f$=85 ohm | Poly | 1.889 | 1 |
| | RBF | 0.250 | 1 |
| RYB fault at 10%, FIA=125° LA=55°, $R_f$=80 ohm | Poly | 1.888 | 1 |
| | RBF | 0.246 | 1 |

Section identification is accomplished by training the $\text{LSSVM}_S$ to build an optimized classifier. The current samples corresponding to one-fourth cycle of post fault current are fed as input to the $\text{LSSVM}_S$ and the output is +1 (before series capacitor) or -1 (after series capacitor). The training is done using 208 data sets and the accuracy of identification is tested over 916 datasets. The average classification rate for section identification for 916 data sets is found to be 97.25 % for all fault types which is reasonably good. Table II depicts the results for section identification with proposed approach. Also, misclassification is observed for the RYB-G fault at 75%, FIA=120°, LA=65°, $R_f$=80ohm, Poly misclassifies showing the output as 1. Other cases shown in Table II shows correct identification.

## IV. SIMULATION STUDIES

A three phase transmission line of 250 km length, 230 kV, 50 Hz connecting two systems with MOV and series capacitor kept at midpoint of line as shown in Fig.3 is used for testing of proposed scheme. The model of transmission system with TCSC shown in Fig.3 has been simulated using MATLAB/Simulink environment. The various parameters of transmission line used for simulation are given in Table III. The post-fault current signal is retrieved at the relaying end bus A and used for recognition of fault events. The sampling rate used is 1.0 kHz at nominal frequency 50 Hz, thus giving 20 samples per cycle.

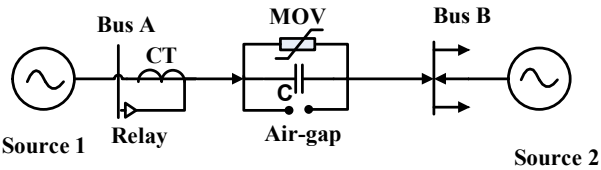

Fig.3 Series compensated transmission line

TABLE III

PARAMETERS OF THE SIMULATED TRANSMISSION LINE

| System voltage | 230 kV |
|---|---|
| System frequency | 50 Hz |
| Voltage of source 1 | 1.0 0 degree pu |
| Voltage of source 2 | 1.0 ∠ 20 degree pu |
| Transmission line length | 250 km |
| Positive sequence | R = 0.0368 (ohm/km), L = 0.55(mH/km), C = 0.028 (uF/km) |
| Zero sequence | R = 0.0328 (ohm/km), L=1.722 (mH/km), C = 0.024 (uF/km) |
| Series compensated | 70 % C =178.54 of |
| MOV | 40 kV 5 MJ |
| Current transformer (CT) | 230kV, 50 Hz, 2000:1 turns ratio |

To demonstrate the potential of the proposed approach, all ten cases of fault event are simulated. The data-set consists of 916 test cases comprising all the ten types of faults with varying fault resistances, different fault inception angles, different source impedance values, different fault positions, different percentage compensation levels for validation of proposed approach. Fig.4 shows the current signals of three phases in case of R-G fault with fault resistance 50 ohm and fault inception angle of 50 degree with 50% compensation level.

## V. RESULTS AND DISCUSSIONS

The classification accuracy of LSSVM-R classifier with RBF kernel as function of two parameters is shown in Fig.6. There is no significant change observed in accuracy as these parameters are varied. The proposed LSSVM based classifier performs consistently well and gives accurate classification for all fault types. This suggests the efficacy of proposed technique. The classification accuracy indicates the average performance obtained by different types of kernel functions used in the proposed scheme. Fig.6 shows overall classification accuracy of the proposed technique which again supports its utility in fault classification task.

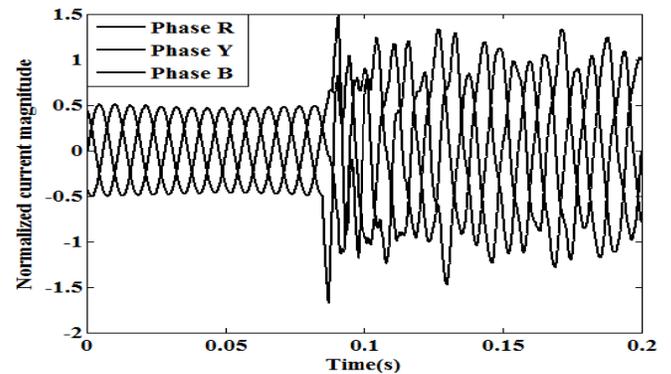

Fig.4 Current signal for the R-G fault

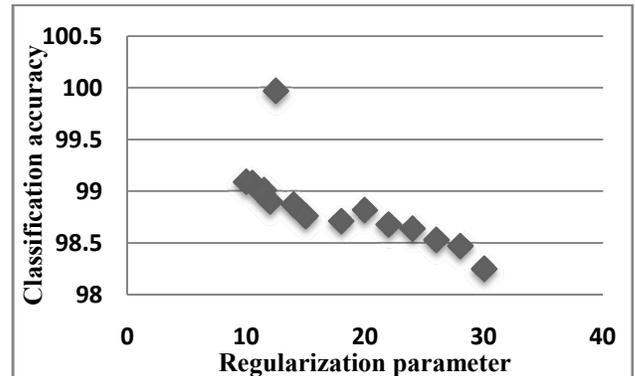

(a) Regularization parameter, $\gamma$

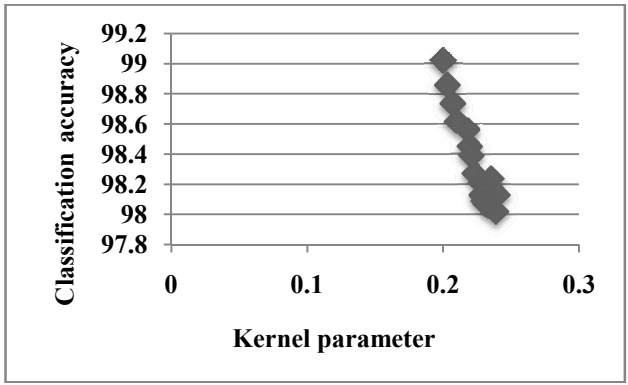

(b) Kernel parameter, $\sigma^2$

Fig.5  Classification accuracy of LSSVM-$_R$ classifier with RBF Kernel

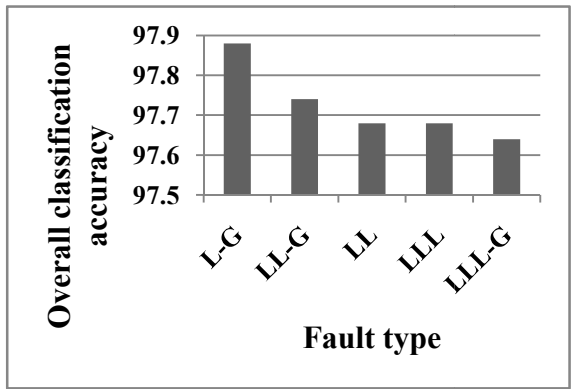

Fig.6 Classification accuracy of the proposed scheme

## VI. CONCLUSION

The present paper investigates a novel approach for fault classification and section identification in a series compensated transmission line. The proposed scheme is based on a modular network consisting of four binary LSSVM classifiers. The proposed classifier has been tested under various system changing conditions. One-fourth cycle of post-fault current samples are used as input by classification scheme. Simulation result validates the consistency and accuracy of proposed scheme. It is found that LSSVMs require lesser training sets for most optimized classification with less number of training samples compared to the neural network and neuro-fuzzy systems. Hence the proposed method is comparatively fast, accurate and robust for the protection of transmission line.

REFERENCES

[1]  A.G.Phadke, J.S.Thorp, "Computer Relaying for Power Systems", Second edition, Wiley-IEEE press, Sept. 2009 (ISBN: 978-0-470-05713-1).
[2]  W.A.Elmore, "Protective Relaying Theory and Applications", 2nd Edition, Marcel Dekker, New York, 2003.
[3]  M.Khederzadeh, T.S.Sidhu, "Impact of TCSC on the protection of transmission lines", IEEE Trans. Power Del., vol. 21,no.1, pp. 80–87, 2006.
[4]  M.Noroozian, L.Angquist, M.Ghandhari, G.Anderson, "Improving power system dynamics by series connected FACTS devices", IEEE Trans. Power Del., vol.12,no.4, pp. 1635–1641, 1997.
[5]  A.A.Girgis, D.G.Hart,"Implementation of Kalman and adaptive Kalman filtering algorithms for digital distance protection on a vector signal processor", IEEE Trans. Power Del., vol. 4, no.1, pp.141-156, Jan.1989.
[6]  Q.Y.Xuan,Y.H.Song, A.T.Johns, R.Morgan, D.Williams, Performance of an adaptive protection scheme for series compensated EHV transmission systems using neural networks. Electric Power Systems Research, vol.36, no.1, pp.57-66,1996.
[7]  Y.H.Song, A.T.Johns, Q.Y.Xuan, "Artificial neural network based protective scheme for controllable series compensated EHV transmission lines", IET Gener. Transm. Distrib., vol.143, no.6, pp.535–540,1996.
[8]  A.Y.Abdelaziz,A.M.Ibrahim, M.Mansour, H.E.Talaat, "Modern approaches for protection of series compensated transmission lines", Electric Power Systems Research, vol.75, no.1, pp.85-98, 2005.
[9]  A.I.Megahed, M.A.Monem, A.E.Bayoumy,"Usage of wavelet transform in the protection of series-compensated transmission lines", IEEE Trans. Power Del.,vol.21,no.3,pp.1213-1221,2006.
[10] P.K.Dash, S.R.Samantray, "Phase selection and fault section identification in thyristor controlled series compensated line using discrete wavelet transform", Electrical Power and Energy Systems vol.26, no.9,pp.725-732, 2004.
[11] M.J.Reddy, D.K.Mohanta, "A wavelet-fuzzy combined approach for classification and location of transmission line faults", Electrical Power and Energy Systems, vol.29, no.9, pp.669-678, 2007.
[12] A.K.Pradhan, A.Routray, D.K.Pradhan, "Wavelet fuzzy combined approach for fault classification of a series-compensated transmission line", IEEE Trans. Power Del., vol.19, no.4, pp.1612-1620, 2004.
[13] A.K.Pradhan, A.Routray, B.Biswal, "Higher order statistics–fuzzy integrated scheme for fault classification of a series compensated line", IEEE Trans Power Del., Vol.19, no.2, pp.891-893, 2004.
[14] P.K.Dash, S.R.Samantaray, G.Panda, "Fault classification and section identification of an advanced series-compensated transmission line using support vector machine", IEEE Trans. Power Del., vol.22, no.1,pp.67–73, 2007.
[15] B.Ravikumar, D.Thukaramand, H.P.Khincha, "Application of support vector machines for fault diagnosis in power transmission system", IET Gener. Transm. Distrib., vol. 2, no.1,pp.119–130, 2008.
[16] U.B.Parikh, B.Das, and R.P.Maheshwari, "Combined wavelet-SVM technique for fault zone detection in a series compensated transmission line", IEEE Trans. Power Del., vol.23, no. 4, pp. 1789-1794, Oct. 2008.
[17] U.B.Parikh, B.Das, and, R.P.Maheshwari, "Fault classification technique for series compensated transmission line using support vector machine", Electrical Power and Energy Systems 32(2010) 629–636.
[18] J.A.K.Suykens, L.Lukas, P.V.Dooren, B.deMoor and J. Vandewalle, "Least square support vector machine classifiers: A large scale algorithm," Neural Process. Lett.,vol.9, pp. 293-300, 1999.
[19] J.A.K.Suykens, T.V.Gestel, J.D.Brabanter, B.DeMoor, and J.Vandewalle, "Least square support vector machine". First edition, World Scientific, Singapore, 2002 (ISBN 981-238-151-1).
[20] Lu BL, and M. Ito, "Task decomposition and module combination based on class relations: a modular neural network for pattern classification," IEEE Trans. on Neural Networks, vol. 10, pp. 1224-1255, 1999.